# The art of molecular computing: whence and whither

Sahana Gangadharan[1,2] and Karthik Raman[1,2,3]

[1]Bhupat and Jyoti Mehta School of Biosciences, Department of Biotechnology, Indian Institute of Technology Madras, Chennai - 600 036, INDIA
[2]Initiative for Biological Systems Engineering, Indian Institute of Technology Madras
[3]Robert Bosch Centre for Data Science and Artificial Intelligence (RBCDSAI), Indian Institute of Technology Madras

## Abstract

An astonishingly diverse biomolecular circuitry orchestrates the functioning machinery underlying every living cell. These biomolecules and their circuits have been engineered not only for various industrial applications but also to perform other atypical functions that they were not evolved for—including computation. Various kinds of computational challenges, such as solving NP-complete problems with many variables, logical computation, neural network operations, and cryptography, have all been attempted through this unconventional computing paradigm. In this review, we highlight key experiments across three different 'eras' of molecular computation, beginning with molecular solutions, transitioning to logic circuits and ultimately, more complex molecular networks. We also discuss a variety of applications of molecular computation, from solving NP-hard problems to self-assembled nanostructures for delivering molecules, and provide a glimpse into the exciting potential that molecular computing holds for the future.

# INTRODUCTION

Silicon-based computers have been mainstream for the last several decades. They have enabled us to solve many computationally demanding and challenging problems. In a world of conventional paradigms and design architectures, biomolecules offer us an unconventional avenue for solving computational problems. Conceivably, some form of molecular computation underlies cellular decision-making, but can we exploit biomolecular processes to solve computational challenges? Simple manipulations to the existing "operating system" of biomolecules provide us with versatile tools and methods that can be leveraged to solve computational problems. Humans have always strived to exploit living cells for various purposes and make the best use of what's already available in nature. This has led to interesting applications that go well beyond natural biological function.

It is interesting to view the history of molecular computing through the lens of *synthetic biology*. Synthetic biology is a multidisciplinary field that involves assembling biological modules distinctively and predictively, to engineer novel circuits and cells for various applications, simultaneously advancing our understanding of fundamental biology. This, as we can imagine, requires breaking down existing networks to their molecular components and assembling them in a new set-up to yield desirable outcomes. Splicing such molecular components for creating biological modules was first proposed by Salvador Luria *et al*[1] and later adapted by Jacob and Monod, for their study of the *lac* operon in *E. coli*[2]. This set the stage for other discoveries such as restriction enzymes and strategies to control gene expression, which are the key methodologies used in this field.

The field of synthetic biology has offered fascinating solutions to various problems. It provides a glimpse into the versatility of cellular networks and how they can be adapted for various applications. These applications include formulating "green chemicals" from agricultural wastes[3], biomimicry[4], rewriting genetic code[5], designing a minimal functioning genome[6], engineering bacteria to replace existing therapeutics[7], and even more interestingly, DNA for data storage[8] and molecular computation[9]. This new-fangled form of computation involves the use of biological elements to solve computational problems.

Unconventional computing approaches may also require unconventional algorithms—the inherent parallelism of DNA and its error-prone nature of replication may require us to develop radically different algorithmic designs to come up with novel solutions for computationally challenging problems. But what kind of molecular computing systems already exist? Are there other computing paradigms that are amenable for solving such problems? Will they require newer ways of devising algorithms? Can we exploit biological molecules for computation—to recall von Neumann[10,11] "*is it possible to design reliable systems from unreliable components*"? Do these systems present important advantages beyond their apparently obvious difficulties and disadvantages? In this review, we seek to answer all these questions, presenting a critical account of several important studies over the last few decades.

The following sections provide a brief overview of the history of molecular computation, starting with Leonard Adleman's seminal experiment in 1994[9]. We delineate the history of molecular computing into three eras: (i) the early history, which pays specific attention to Adleman's classic work and other 'molecular solutions' to computational problems, followed by (ii) the era of logic circuits, where the emphasis moved to logic circuits and nano-assembly, and finally, (iii) the era of molecular networks, where DNA neural nets, self-assembly-based DNA robots and information processing techniques are emerging as state-of-the-art technologies. Our article will majorly focus on the methodologies and applications developed in DNA computation and briefly touch upon peptide and cellular computation. We systematically map out existing research using a generalised morphological analysis (GMA), illuminating important connections between molecular computing approaches, the nature of computationally solved problem, and the molecular techniques used, to ultimately point out potential gaps that future research can address. We discuss the key applications developed in this field, as well as the promise and pitfalls of molecular computation.

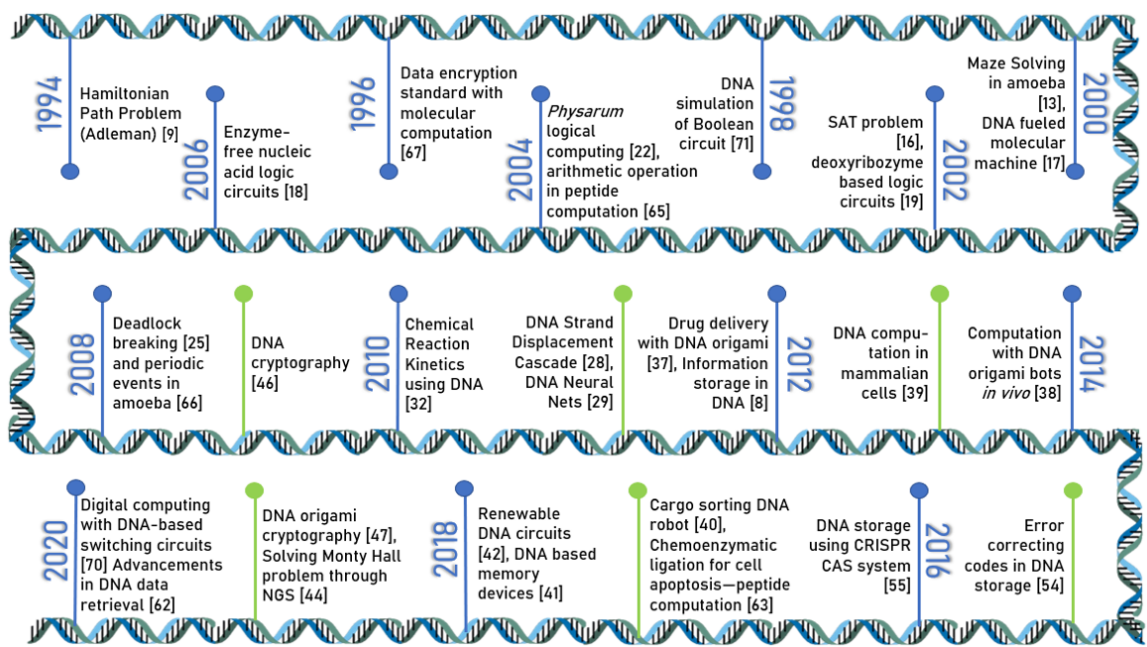

**Figure 1: Timeline of major breakthroughs in the field of molecular computation.** Starting from Adleman's experiment in 1994, the field has come a long way, paving the path for exciting techniques such as DNA Strand Displacement Cascades, DNA self-assembly, nano-robotics with molecules, and neural network computation being the state-of-the-art methodologies today. The numbers in square brackets denote references.

# EARLY HISTORY

Although several studies had discussed the potential of biomolecules for computing and self-assembled structures, it was not until Adleman's seminal experiment in 1994 that the field gained due recognition.

## Solving the Hamiltonian: Adleman's experiment

In 1994, Leonard Adleman provided a proof-of-concept to solve the Hamiltonian path problem[1] using DNA strands. He used an unusual approach to solve this NP-complete problem[2]. He formulated an algorithm that incorporated DNA strands to represent the input graph and Watson–Crick pairing to capture other problem constraints. This landmark experiment laid the foundation for other research and fuelled the growth of this field.

### The algorithm

The algorithm for this problem, as stated by Adleman, is somewhat brute force, yet elegant. The first step is to generate *all* possible paths given a directed network. Next, the paths that begin and end with the specified start and end node are picked. The next step is to identify those paths where all the nodes are visited at least once. Further, it must be ensured that each vertex in the network is visited exactly once. If there remain solutions that satisfy all the above criteria, such paths would be solutions for the Hamiltonian path problem of the given network. This simple and intuitive algorithm can be implemented using a system in which oligonucleotides represent the nodes and the edges of the given network. The experimental strategies are truly novel, making this one of the most exceptional studies.

### "DNA implementation" of the algorithm

The implementation of the algorithm was experimentally demonstrated on a 7-node directed graph. The nodes (numbered 0 through 6) were represented by random 20-mer sequences of DNA, and the edges were designed such that they shared sequences with both the vertices that connected them. A ligation reaction, followed by a Polymerase Chain Reaction (PCR) with the 20-mer sequence of node 0 and 20-mer complementary sequence of node 6 as primers, was set up. This generated random paths with the appropriate starting and ending nodes. Once the resulting DNA strands were run on an agarose gel, the portion that corresponded to the desired length of the solution path (140 base pairs), was lysed from the gel. The product from the above step was further affinity-purified to make sure that all the vertices of the network were present in the solution-path exactly once. Finally, a graduated PCR was run on the product to yield a higher concentration of that particular DNA strand—the solution for the Hamiltonian Path Problem of the chosen network.

---

[1]A classic NP-complete computational problem, involving finding a Hamiltonian path in a graph—a path that passes through every vertex in the graph exactly once; it also bears an interesting relation to the travelling salesperson problem.

[2]NP-complete are those problems which do not have a "polynomial time solution", i.e. the time taken to solve the problem typically increases exponentially with problem size. But, if given a solution, it is easy to verify that the solution is correct, in polynomial time.

Of course, merely solving a 7-node Hamiltonian problem was not the major contribution of this work; rather, it heralded a new paradigm for computation, using biomolecules. The massive parallelization involved in generating all paths underlines the power of this approach, given that $10^{14}$ operations were executed just by the end of the first step. The method was also remarkably energy-efficient, requiring barely 1J of energy to run $2 \times 10^{19}$ operations.

## The dawn of other forms of molecular computation

Beyond DNA, other biomolecules have also been exploited to solve computational problems. The following studies use RNA and a cellular system to solve satisfiability and travelling salesperson problems, respectively.

### Molecular computation: RNA solutions to chess problems

Landweber and co-workers[12] proposed an algorithm to solve a modification of the Knights problem, which asks for the configuration of knights on an $n \times n$ chessboard such that no two knights can attack each other, with the help of RNase H Digestion. A $3 \times 3$ chessboard, where RNA oligonucleotides represented each position, was employed for the experimental procedure. The key idea was that the RNase would hydrolyse RNA strands that did not fit the constraints of the problem. The RNA strands were chosen under the basis of certain principles, and the hydrolyzation reaction was done iteratively, according to the logical encoding of the problem. The paper suggests that molecular computation is able to find even the rarest of rare solutions, for instance, even a single strand from over a quadrillion, for a combinatorial problem.

### Maze-solving by an amoeboid organism

*Physarum polycephalum* is a slime mould that has dendritic networks of pseudopodia. It was observed that when this mould was placed at the beginning of a small labyrinth, with nutrient agar at both the start and the end, it always preferred the shortest path to its food source[13]. Instead of spreading out the pseudopodia into different routes, the mould seemed to swirl itself into one thick tube, thereby increasing its foraging efficiency as well as its chance of survival. Thus, it appeared to display the primitive intelligence required to solve an otherwise mathematically hard problem — the Travelling Salesperson Problem (TSP). Here, the cell as a whole indulges in the operation performed thereby introducing cellular computation, rather than biomolecular computation.

# THE ERA OF LOGIC CIRCUITS

Following the development of molecular solutions to computational problems, the first decade of the 21st century (2000-2010) saw a number of experiments, mostly rooted in logic gates and arithmetic computations, leading to rapid growth in the field.

## Logical Computing with DNA molecules

Adleman's solution to a major combinatorial problem motivated the rest of the scientific community to try out other such problems. One interesting puzzle from that class of problems is the **3-SAT satisfiability problem**[3] which is NP-complete. Several versions of the problem have been solved through diverse approaches. For instance, Smith[14] solved the 3-SAT problem with four variables, analysing $2^4 = 16$ truth assignments, and Sakamoto *et al*[15] solved a six-variable problem, finding the right solution amidst $2^6 = 64$ possible assignments. But a few years down the line, Braich *et al*[16] proposed to solve the problem with 20 variables, thereby scanning through $2^{20}$ possible truth assignments. Also, to increase the difficulty, the problem was set such that there was only one correct answer in the entire search space.

For the solution, oligonucleotides representing variables were synthesised on a solid support using a mix-and-split combinatorial technique, thereby allowing the computation to happen on its surface. The constructed device had two chambers, with an acrydite-modified probe in each of them. The strands holding the correct truth assignments were captured in the cold chamber, and the rest were released from the hot chamber. This way, the strand representing the factual truth assignments of all the 20 variables was secured at the cold chamber. For the most part, the sole "operation" in this experiment was based only on Watson–Crick pairing and melting of DNA. The implementation of logic gates using DNA was the next breakthrough in this field.

The first portrayal of a DNA circuit was proposed by Yurke *et al*[17]. In this study, they showed two sets of DNA strands that precisely aided one another in nano-scale movements. The first set of strands constituted the circuit, while the second set acted as a nanomechanical switch that fuelled the movement of the first set. Nucleic acid-based logic circuits were later introduced by Seelig *et al*[18] where the designing and modular construction of several logic gates based on miRNA and DNA sequence recognition are discussed, which includes features such as amplification, cascading and feedback loops. Stojanovic *et al* proposed a **deoxyribozyme-based modular design**[19] that was used to generate any Boolean function.

---

[3] Satisfiability problem is a problem of determining if there exist replacements of variables in a specific Boolean expression such that the formula evaluates to TRUE.

Stojanovic and Stefanovic later improvised on their initial work by devising a molecular automaton[20]. This formulation was designed to play tic-tac-toe with a human opponent, and always guaranteed either a win or a draw. The Molecular Array of YES and ANDANDNOT gates was named MAYA. The human's turn in the play corresponds to the addition of a new DNA strand to which the automaton will initiate a response which could be observed by reading the fluorescence effect from each compartment. MAYA, unlike the methodologies assumed by previously mentioned experiments, works uniquely, by implementing a digital logic circuit using enzymes from metabolic circuits.

## Peptide strands for solving mathematically hard problems

Peptide strands are made up of 20 variant building blocks and this confers them advantages over DNA strands for unconventional computation, as the density of information that can be stored in a protein sequence is relatively high. Of course, the kind of tools and techniques required to "operate" on peptide data would differ; the central tool happens to be the specific interaction between epitopes and monoclonal antibodies. Hug and Schuler[21] attempted various combinatorial problems, such as comparing the frequency of a single element in different sets, estimating the number of times an element occurs in a given set and satisfiability problems. However, other solutions provided by DNA strands were not easily replicable using peptide computation, since there were only a specific set of epitopes–monoclonal antibody pairs that were discovered then, limiting the complexity of problems that could be attempted.

## Logical Computing with cells

Beyond molecules, entire cells have been exploited for computation. Tsuda *et al*[22] discuss the idea of logical computing in *Physarum* and what happens when there is hardware damage in the system. The logical operations were based on chemotaxis, aversion phenomenon of *Physarum* against each other and their ability to fuse under extreme situations. AND, OR, and NOT gates were built based on the established guidelines, and expected results were observed. A part of the AND gate was broken to test the emergent behaviour of *Physarum.* In such a scenario, the expected outcome for 1 AND 1 was 0, but on the contrary, the observed results were quite the opposite. One *Physarum* fragment, instead of averting itself from the other fragment, chose to fuse with it and they moved towards the right output path together, leading to 1AND1 = 1. Such inherent behaviour of the *Physarum* leading to self-repair upon damage shows us the possibility of using living organisms to solve logical problems. Jones and Adamatzky[23] computationally designed **half-adder and full-adder circuits** using *Physarum,* thereby demonstrating the organism's ability to adapt to any kind of environment and emphasising its robustness.

An interesting application of *Physarum*[24] was a **bio-robot** whose only source of instructions came from the physical environment. The external cues were coupled with the intracellular information processing ability of *Physarum* to yield an omni-directional walker that could solve optimisation problems.

*Physarum* was also observed to display properties that enabled it to flexibly **break deadlock-like situations**[25]. There is no higher-level authoritative controller in these systems, thereby making its sub-components behave independently while still maintaining favourable conditions for the whole system. In a sense, *Physarum* implements a recurrent neural network algorithm which enables the amoeba to solve a constrained satisfaction problem. The amoeba is captured in a star-like structure so that any movement in the amoeba makes it traverse in multiple branches. These branches are photosensitive and operate concurrently as processors of the system. An optical feedback control is introduced to create conflicts within the processors, thereby allowing them to function as a recurrent neural network for imposing a particular constraint satisfaction problem.

# THE ERA OF MOLECULAR NETWORKS

## Digital circuits and Origami with DNA strands

This era began with many striking advances by Erik Winfree and Lulu Qian. They introduce this technique where, metaphorically, **DNA strands play on a See-Saw**[26]. Two single strands of DNA that have a complementary sequence to a template strand anneal and melt from the template strand in a see-saw fashion, thereby making a wide range of applications possible. This mechanism, called the *'toehold-mediated DNA strand displacement'*, was originally proposed by Yurke *et al*[17], and later popularised by Qian and Winfree. This design was used to construct other circuits such as feedforward digital circuits, relay contact circuits and analogue time-domain circuits[27].

Further, they extended the idea to develop logic gates. A specific arrangement of OR and AND gates also permitted the construction of a 3-bit XOR gate and the computation of square root of a 4-bit number. As the complexity of the problem at hand increased, they set up a debugging tool which was another output 'wire', from an intermediate step, through which the functioning of the entire circuit could be analysed. It is easy to envisage the construction of more complex circuitry from these standard building blocks, which underlie all electronic circuits. This way, they furthered the notion of digital computation using molecular circuits[28].

Next, they built a **neural network using DNA sequences**[29], where they assembled a sophisticated circuit, involving 72 DNA species, which resembled a Hopfield network. A Hopfield network[30], is made up of artificial neurons that are fully connected and possess the ability to remember a set of patterns from their training set. The functionality of this molecular network was tested in a cuvette via a game called "read your mind". In this game, a human would select and keep one of four given scientist's names in his/her mind and would answer a set of questions, by giving inputs to the associative memory network. Based on the answers to the predefined questions, the network was able to correctly guess the scientist, sometimes even before the opponent gave out all the available hints by answering the questions. The authors showed that the network can handle robust inputs and any other defect within the system, including sparse connectivity of the network. The 'Winner-take-all' algorithm by Cherry and Qian[31] employed the above algorithm to the MNIST problem[4] using DNA strands that worked on the basis of their ability to remember patterns and classify complex and noisy data. The results of their work have paved the path for biomolecular systems that can learn, memorise and recall.

Toe-hold strand displacement cascades can also be employed in several other aspects. The intrinsic analogue nature of biomolecules enables a much broader class of behaviours. In particular, **Chemical Reaction Networks** (CRNs), introduced by Soloveichik *et al*[32] operate as a 'programming language' that facilitates the study of the underlying kinetics in unimolecular and bimolecular reactions, while also discussing limitations and challenges. CRNs can be thought of as a general framework for modelling networks with interacting species. Nano-controllers were implemented by Chen *et al*[33] using plasmid DNA to test and study fundamental reaction types, eventually leading to *de novo* engineering of interactions between the designed components.

In early 1982, Seeman[34] explained the different types of junctions and lattices that can be formed using nucleic acids. Seeman elucidates the basic rules that must be followed to generate covalently bond lattices that are periodic in connectivity and plausibly in space too. DNA-based tile-assembly is a versatile toolbox that provides directions for nanoarchitecture and directed self-assembly of DNA strands[35]. It facilitates structures with increased complexity and behaves as a scaffold that aids in the assembly. Tile-based work also supports robust reprogrammable computations[36] to execute different algorithms, thereby welcoming molecular engineering to the algorithmic era. Jiang *et al*[37] described an unconventional method for drug delivery, where they used **DNA origami-based nanostructures** that could self-assemble and more importantly, were biocompatible and spatially addressable. Amidst all the exciting *in vitro* and *in silico* research, Amir *et al*[38] fabricated a DNA nanobot that could control a specific molecule *in vivo*, inside a living cockroach. Another such circuit that responded to specific miRNAs in the cellular space was constructed, to enable the monitoring and imaging of cell-specific markers[39].

A DNA-based nanobot was designed for sorting two types of biomolecules into separate piles[40]. The bot would perform a random walk across a two-dimensional DNA origami surface, thereby saving energy, picking up and dropping biomolecules as and when it encountered them. DNA that is biocompatible can thus be used to record specific intracellular events that might be very significant in diagnostics and treatment. The methodologies for recording and retrieving "such events" were reviewed by Sheth and Wang[41]. A target DNA sequence is altered based on a set of rules depending on the spatio-temporal localisation of the intracellular event. This can happen at a single directed location, or at multiple target locations in either a targeted manner or by accepting stochastic changes in the DNA sequence. The altered DNA is then sequenced and analysed to identify the type and rate of changes that happened in the intracellular environment to infer specific details that may be very difficult to observe in the absence of this technique.

---

[4] MNIST is a large database of handwritten digits, and a gold standard dataset for testing image processing systems.

Mamet *et al*[44] showcased the solution for the "Monty Hall problem" with DNA strands. The authors used Illumina Next Generation Sequencing to identify the strand that corresponds to the correct answer over multiple simulations. Experimentation and retrieval of results are usually time-consuming; yet, this task was done within 24 hours, including the preparatory work.

## DNA Cryptography

Storing information in DNA also inspired the idea of storing and retrieving cryptic messages inside the biomolecule, giving rise to **DNA Cryptography**. This idea was first proposed in 1995[45], and since then, many approaches have been tried and tested. A 'cryptographic' experiment was attempted by Ning[46], in which he encoded data in a DNA sequence and allowed it to undergo the steps of the Central Dogma, with introns being spliced in the process. To transfer the information as a secret message, the receiver is given the peptide sequence through a public channel, the additional information about the regions that underwent splicing and a cryptic message regarding the introns is shared through a secure channel. This way, even if the peptide sequence is obtained by an intruder, they will not have the complete information required to decode the original message. Another recent DNA cryptographic experiment involves self-assembling origami structures[47]. The secret message is first encoded as a braille-like pattern in the outer layer. This is followed by the formation of a steganographic intermediate layer. The receiver, upon receiving these scaffolds, uses specific staple sequences to fold the origami, revealing the original patterns under an atomic force microscope. The authors suggested DNA origami cryptography as a biomolecular solution that meets the modern demands of high information security.

## Exploring the field through competitions

Various synbio competitions have served to further increase the excitement around molecular computation. The **International Genetically Engineered Machine** (*iGEM: http://www.igem.org/*) competition[48], is a worldwide synthetic biology competition which promotes a systematic study of biology and a platform to develop new ideas and test them with support and guidance from the experts. Another initiative is the **BIOMOD competition** *(http://biomod.net/)*, led by Shawn Douglas, where numerous students participate in designing devices with biomolecules for various applications. This competition has brought forth some of the most exquisite works in the field of molecular nanotechnology and computation. The **Molecular Programming Project** *(MPP; http://molecular-programming.org/)* is an initiative by 11 scientists from various departments at Caltech, Harvard, University of Washington and University of California, San Francisco. This initiative aims to incorporate computer science and programming principles into molecular systems, thereby creating nanoscale devices that have potential applications in robotics and the biomedical industry. Further information on these competitions and initiatives are provided in the Supplementary Table 1. They have served to accelerate growth in these fields, also creating many opportunities for the future. Most importantly, they have enthused and attracted many young researchers to dabble in this area.

DNA for Data Storage

An unconventional application of DNA is in **information storage**. DNA arranges itself in a very compact manner, and its miniature size supports extremely dense information storage, with an estimated density of 5.5 Pb/mm[8]. In comparison, a microSDXC card[49] has a density that is approximately a million times lower[5]. Although storing data in DNA and retrieval are tedious, DNA can store large amounts of data for a long time. This technique was first implemented in 1988, by J. Davis[50], when he successfully cloned a picture of *Microvenus,* encoded in DNA *bases*, onto *E. coli*. Since then, many attempts have been made at storing different kinds of data in DNA. In 2012, George Church and his team encoded 70 copies of an HTML version of his book, *Regenesis* [51], as DNA sequences[8]. Following this work, a team led by Nick Goldman encoded all 154 sonnets by William Shakespeare, an audio clip from Martin Luther King's famous speech, *"I have a dream"*, the classic paper on the structure of DNA by James Watson and Francis Crick, a picture of the researcher's institute, followed by the instructions on how the data can be converted into a single DNA sequence[52]. The properties of DNA as a storage device and its efficiency over other contemporary storage media are discussed by Zhirnov *et al*[53]. Studies have also looked at error-correcting codes[54] and *in vivo* storage using CRISPR Cas systems[55]. An excellent review of DNA data storage has been published elsewhere[56].

Although there are obvious advantages of this technique, it is important to note the limitations as well. The chemical synthesis of DNA is still costly[57] and the mechanism for identification and the retrieval of specific messages is still tricky[58]. However, in the recent past, there has been significant research resulting in the development of appropriate methodologies and tools for accessing and retrieving data from DNA. Yazdi *et al*[59] proposed PCR-based random access of data while also offering methods for re-writability. Random access over large data was first suggested by Karin Strauss and co-workers[60]. Further, Microsoft demonstrated a fully automated process for encoding and retrieving messages[61]. The reliable retrieval of messages[62] when very few copies of the message are present in the DNA pool, was shown by yet another team from Microsoft. These advances indicate the extensive potential of DNA as "wetware" with its resilience as an added advantage.

---

[5] The maximum storage of the latest microSDXC cards is 1TB (=8000 Gb); and the dimensions of the card are 32 x 24 x 2.1-1.4 mm. Therefore, the information density of this device is - 8000/ (32*24*2.1) - 8000/ (32*24*1.4) = 4.96 - 7.44 Gb/mm³, which is ~5–7 Gb/mm³.

## Peptide computation

Molecular devices with enzymes as processors were adapted to perform logical computing with peptide sequences as inputs and apoptosis for cancer cells as the output. Logic operations such as AND (where the presence of both input molecules was essential to initiate the reaction), OR (where the presence of one of the input peptides would suffice to initiate apoptosis) and INHIBIT (where the presence of only one input would initiate cell death) were performed using suitable peptide sequences. Following this, a 3-bit peptide keypad based on concatenated logic gates was designed. This device only works when appropriate sequences are input, especially in the correct order. When this was tested with normal and HeLa cells, apoptosis was observed only in the latter even when the right set of inputs was given to the former. This demonstration by Li *et al*[63] seems to be extremely beneficial for applications such as targeted cell apoptosis and cell differentiation.

# KEY APPLICATIONS DEVELOPED

The range and scope of molecular computation have expanded extensively, and have come a long way since Adleman's first experiment. We here present a GMA of the literature, capturing important connections between molecular computing approaches across the eras, the nature of computational problems that have been solved, as well as the molecular techniques used[64]. In Table 1, we have listed the functionalities and the approaches taken in biomolecular computing, along the rows, and the different types of problems encountered, along the columns. A cross-consistency matrix is built by placing all the papers referred to in this article, in the appropriate boxes. This matrix systematically maps out literature enabling the identification of unique connections that are otherwise imperceivable to this extent, through conventional methods. The gaps in the matrix suggest either a lack of research in the area, or an error in the literature survey. However, to the best of our knowledge, we have not left out any major contributions in the field. For instance, we speculate that there have not been works employing self-assembly to solve RNA-mediated chemical reactions kinetics. The GMA is thus particularly useful to identify less-researched areas that offers potential research questions to work upon.

Starting from NP-Complete and Combinatorial problems such as the 3-SAT satisfiability problem and TSP, DNA computation has aided in solving diverse sets of problems, including that of logical and neural network computation. By the beginning of the molecular networks era, a DNA computing device could perform the basic functions of a contemporary silicon-based system. Complex designs such as the implementation of neural networks using DNA sequences, nanobots that could sort different kinds of molecules, DNA architecture that could track and image molecules, both *in vitro* and *ex vivo,* were demonstrated. Renewable circuits that allow the use of the same device for multiple computations, Illumina sequencing that facilitate parallel processing and of course the use of DNA as a medium for storing data and cryptic messages contributed to further advances in the field. The implementations of these problems were predominantly based on Watson–Crick complementary base-pairing, annealing features, deoxyribozyme-based methods, and DNA strand displacement cascades. Advancement in the field of cellular computing paved the pathway for intriguing features such as emergent behaviour, modular design and also the adaptation of chemotaxis for information processing. The growth of peptide computing was also remarkable for its aspects of providing solutions for arithmetic problems, logical computation and finally, the use of peptides as processors[65].

## Advantages and Disadvantages

Molecular computing has several advantages. Firstly, the ability to perform a massive number of reactions concurrently makes molecular systems efficient, allowing the device to perform an exhaustive search within hours of running a PCR reaction. Secondly, their miniature architecture also facilitates their application in tracking molecules inside certain living organisms, benefiting the biomedical community. Finally, the energy efficiency of these systems demonstrates their resourcefulness, outweighing any other current-day computers. Despite the continued validity of Moore's law, and the concomitantly increasing computational power, the last decade has seen an intense focus on power-efficient computing, focusing less on mere teraflops and petaflops, but on per-Watt performance. Molecular computing, despite its inherent difficulties, outshines traditional computing in energy efficiency.

However, it is necessary to analyse the various drawbacks of these systems as well. First, the functioning of these systems involves many steps, starting from encoding the data carefully, performing the prerequisites, to finally sequencing the biomolecules to read out the solution, resulting in a very tedious process. What can be presently solved on a desktop computer in microseconds might require a few hours, or even days, with molecular computing devices. Second, the inherent noise in biological systems makes them unpredictable in certain scenarios. The field is still at its inception, and issues like reproducibility and reliability remain to be addressed.

**Table 1:** An overview of the variety of molecular computation approaches that have been used to tackle various problems.

| | Approach/ Functionality | Combinatorial | Decision making | Mathematical | Chemical Kinetics | Logical Modelling | Memory-based | Information processing and Cryptography |
|---|---|---|---|---|---|---|---|---|
| DNA Computation | Recursive | [9] [14] [15] [16] | | | | | | |
| | High-throughput DNA Sequencing | | [44] | | | | | |
| | RNA mediated | | [44] | | | | | [46][61][62] |
| | Toehold mediated strand displacement | | [12] | | | [39] | | [46] |
| | Self-assembly/ Origami | | | [28][31] | [32][33] | [27][31][39] [68-70] | [29] | |
| | Logical Operations | | [12] [20] | [28][31] | | [35-40] | | [47] |
| | Cellular Recording | | | | | [17] [19-20] [27][31] [67-69] | [29] | [67] |
| | Biosensing | | [12] | [28] | [33] | | [41] | |
| | Renewable devices | | | | | [42][43] | [41] | [47][55] |
| | Storage device | | [54] | | | | | [59] |
| | | | | | | | | [8] [45-47] [52-55][59-62] |
| Peptide Computation | Differential-affinity based | [21] | [21] | [65] | [65] | | | |
| | Biosensing | [21] | [21] | [65] | [65] | | | |
| | Logic Operations | | [63] | | | | | [63] |
| | Cell Regulation | | [63] | | | | | [63] |
| Cellular computation | Spatiotemporal oscillations/ shuttle streaming | [13] | [13] [25] | | | [24] | [66] | [24] [66] |
| | Optical Feedback control | | [25] | | | | | |
| | Logic Operations | | [22] [25] | [23] | | [22] [22] [25] | | |
| | Light-sensing | | | | | [24] | | [24] |

# WHITHER?

Having discussed the the various facets of molecular computation, it is vital to analyse where the field is headed. Applications that employ DNA self-assembly in therapeutics and in the molecular manufacturing sector will certainly benefit the most out of this field. In fact, tackling medical problems with molecular computation has already been taken up by competitions such as iGEM and BIOMOD, (see Supplementary Table 1). We might also expect applications such as CAD designing for DNA, which the MPP is currently attempting, and automated systems that perform operations based on certain rules. These enhancements will reduce human intervention in the procedures, thereby facilitating easier design and mitigating errors, if any. Automated systems are also immensely helpful in regulating design principles and reducing the noise in the system. With substantial investment from technology giants such as Microsoft, DNA data storage seems to be on an accelerated growth trajectory, with automated systems performing the assigned methodologies.

# CONCLUSION

The potential of molecular computation is immense. Employing biomolecules for computation has given rise to various interesting applications from simple arithmetic to logical operations and, finally, to neural networks in which the biomolecular device has the ability to memorise and recall events. The idea of engineering biomolecular circuits to achieve our goals started with Jacob and Monod but has come a long way since. Although biomolecular devices are not as effective as silicon-based systems for basic operations, they have proven to be extremely advantageous for specific use cases and applications. Adleman and others, as we have seen, have beautifully adapted biomolecules and their interactions, advancing the art and theory of biomolecular computation.

To summarize, this field isn't just another molecular biology experiment. Many potential applications await, making the future exciting and inspiring for the community to witness and learn. Additionally, it is important to appreciate that a remarkable variety of complex tasks are carried out by the "cells" in every living being, including sophisticated information processing, and complex decision-making to sustain life and evolve. Any work that taps into these naturally available systems would ultimately enhance our ability to understand, exploit and manipulate these building blocks, providing us with a versatile toolbox for solving a variety of challenges confronting humanity.

# ACKNOWLEDGEMENTS

The authors thank Hariharan Srikrishnan, Sankalpa Venkatraghavan and Sathvik Ananthakrishnan for critical reading of the manuscript and constructive suggestions. We apologise to colleagues whose work we could not cite due to space constraints.

## CONFLICT OF INTEREST

The authors declare no conflict of interest.

# SUPPORTING INFORMATION

# SI Figure 1 - Molecular Tools used in biomolecular computation

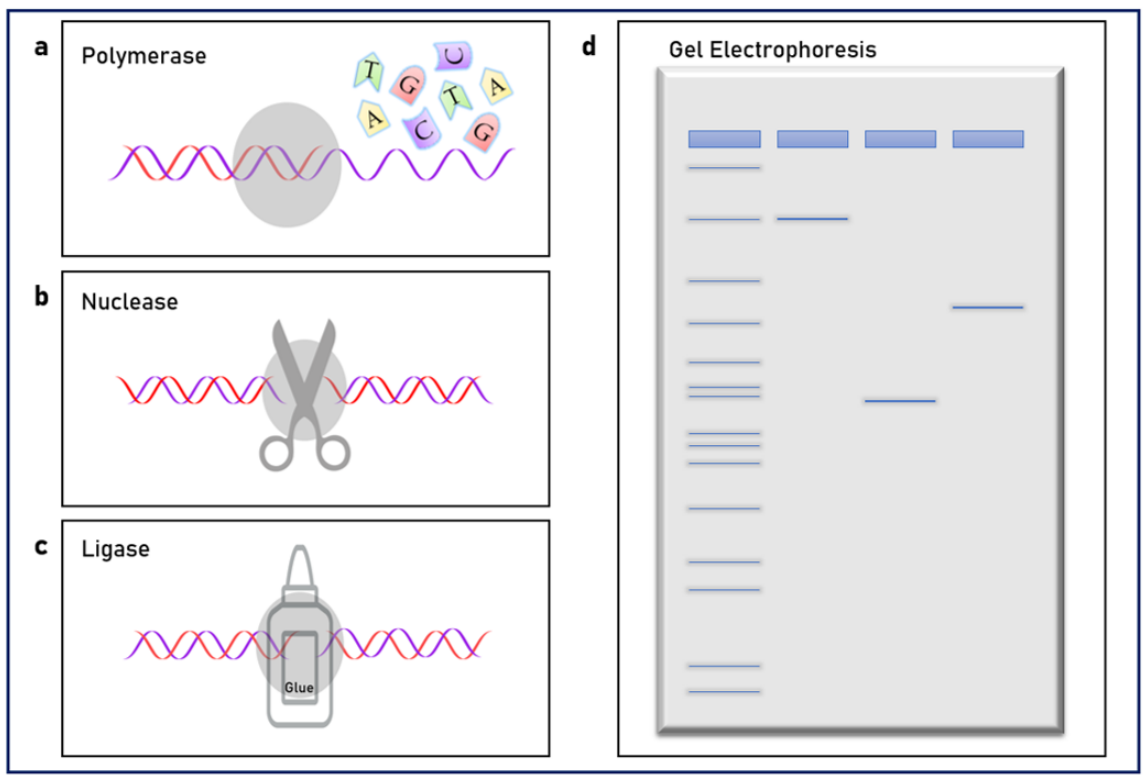

**Fig 1: The major 'tools' of DNA computation. a)** The polymerase is an enzyme that synthesizes DNA strands, given a template strand. The copying mechanism is based on Watson–Crick base-pairing rules. **b)** Nuclease is an enzyme that nicks a double-stranded DNA molecule based on a specific sequence in a strand, therefore acting as a specific pair of scissors. **c)** Ligase is an enzyme that catalyses the joining of two DNA strands, thereby acting as a specific molecular glue. **d)** Gel electrophoresis is a technique used to analyse the length of a DNA strand. As DNA is negatively charged, it gradually moves from the cathode to the anode, upon the passage of an electric field, with smaller strands travelling faster than longer strands. Length is observed by comparing the final positions of the strands in the gel against a labelled DNA ladder.

# SI Figure 2 - Overview of the techniques in the field of DNA computation

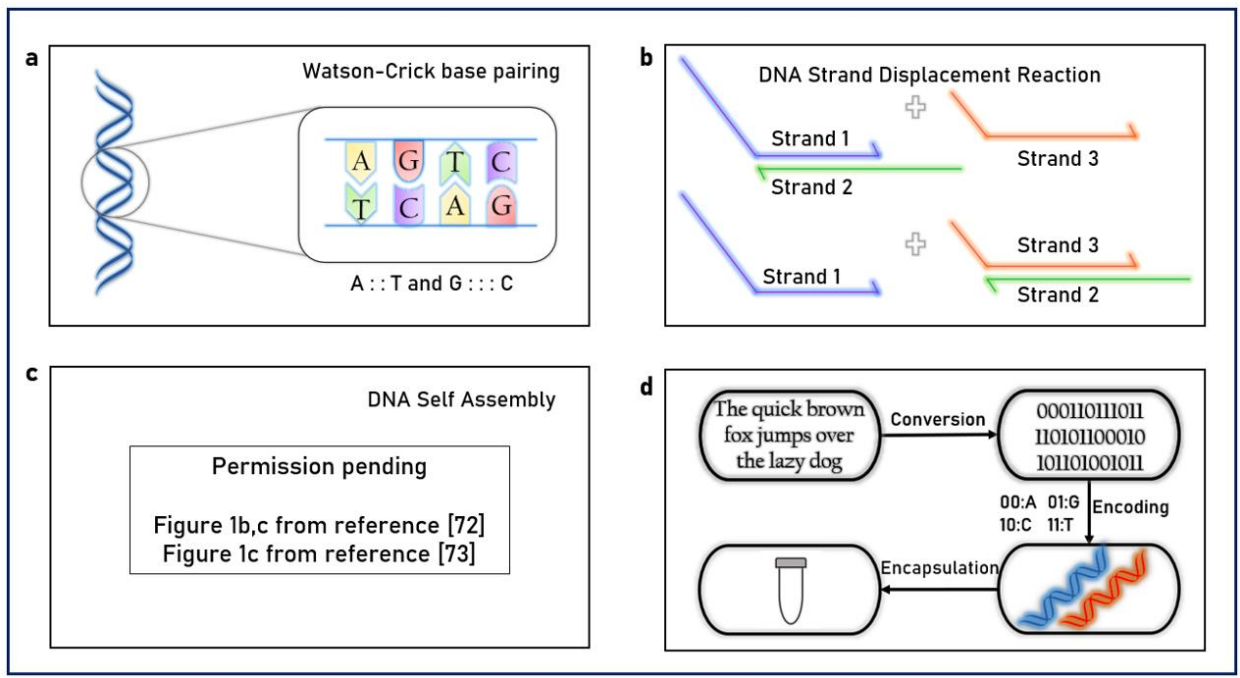

**Fig 2: Overview of the techniques in the field of DNA computation. a)** Watson–Crick complementary base pairing is the one rule that binds all the other techniques as well. It states that the nucleotide Adenine (A) always pairs with Thymine (T) via a double hydrogen bond and nucleotide Cytosine (C) always pairs with Guanine (G) via a triple hydrogen bond. **b)** DNA Strand Displacement Reaction or Toehold-mediated strand displacement is a molecular technique to exchange one strand of DNA with another by employing the process of branch migration. **c)** DNA self-assembly/origami is a very famous technique in which DNA strands fold themselves to form a 3-dimensional structure that performs specific functions. This technique has been used for several proof-of-concept and therapeutic purposes. (*Adapted from references 72 (Fig 1 b, c), 73 (Fig 1c) – permission pending*) **d)** DNA for data storage is a recently developed technique where the biomolecule encodes data. This mechanism offers large storage space and safer ways of storing important data. It is done in 3 steps, namely 1. Conversion (where data is converted into binary bits), 2. Encoding (The binary digits are encoded as DNA strands) and 3. Encapsulation (The DNA strands are further modified to escape degradation and are stored in Ep' tubes).

SI Table 1 - Recent progress in the field of molecular computation fuelled by competitions and initiatives.

| Competition/ Initiative | Team/ Project | URL | Description |
|---|---|---|---|
| iGEM | Team AHUT_China | http://2016.igem.org/Team:AHUT_China | Bio-navigational system that enables a faster solution for the shortest past, given any network |
| iGEM | Team Thessaloniki | https://2019.igem.org/Team:Thessaloniki/Description | DNA computer to recognise specific DNA-protein interactions |
| iGEM | Team SEU (Southeast University) | https://2019.igem.org/Team:SEU | Neural network computation using DNA strands that perform ReLu and Sigmoid functions. Open-source software tools that design DNA reactions. |
| BIOMOD | Team USYD | https://usyd-biomod-2019.webflow.io/project | Cure for Cardiovascular diseases using DNA nanostructures that selectively capture LDL. |
| BIOMOD | Team TU Berlin | http://biomod.biocat.tu-berlin.de/ | Multibrane, a biological membrane with enzyme-rich flagella, that degrades pollutants and heavy metals in water. |
| MPP | NUPACK | http://www.nupack.org/ | Package for analysis and design of nucleic acid structures. |
| MPP | gro | http://depts.washington.edu/soslab/gro/ | Cell programming language that facilitates modelling the behaviour of cells in communities |
| MPP | caDNAno | https://cadnano.org/ | Aids in CAD design of DNA nanostructures |
| MPP | Single-stranded RNA nanostructure | https://science.sciencemag.org/content/345/6198/799 | RNA origami folding where lattices are made by annealing and/or co-transcriptional folding. |